\documentclass{egpubl_arxiv}

\ConferenceSubmission %
 \electronicVersion %

\ifpdf \usepackage[pdftex]{graphicx} \pdfcompresslevel=9
\else \usepackage[dvips]{graphicx} \fi

\PrintedOrElectronic 

\usepackage{t1enc,dfadobe}

\usepackage{egweblnk}
\usepackage{cite}

\usepackage[table]{xcolor}
\definecolor{headercolor}{HTML}{b2df8a}
\definecolor{header2base}{HTML}{AE2573}
\colorlet{header2color}{header2base!30}

\usepackage{booktabs}
\usepackage{graphicx}
\usepackage{tabularx}

\title[PointGrade: Geometric Priors for Grading MoonBoard Problems]%
      {PointGrade: Geometric Priors for Grading MoonBoard Problems}

\author[Stotz et al.]
{
\parbox{\textwidth}{\vspace{-1.5cm}\centering B. Stotz$^{1}$, N. Wang$^{1}$, D. Nogina$^{1}$, C. Zhang$^{1}$, J. Yin$^{1}$, A. Huang$^{1}$, B. Yang$^{1}$, J. Li$^{1}$, J. Salzman$^{1,2}$, S. Feiner$^{1}$, S. Sell\'an$^{1}$
        }
        \\
{\parbox{\textwidth}{\vspace{-1.6cm}\centering $^1$ Columbia University\qquad
        $^2$ Brown University
       }
}
}

\begin{document}

\teaser{
\vspace{-1.9cm}
 \includegraphics{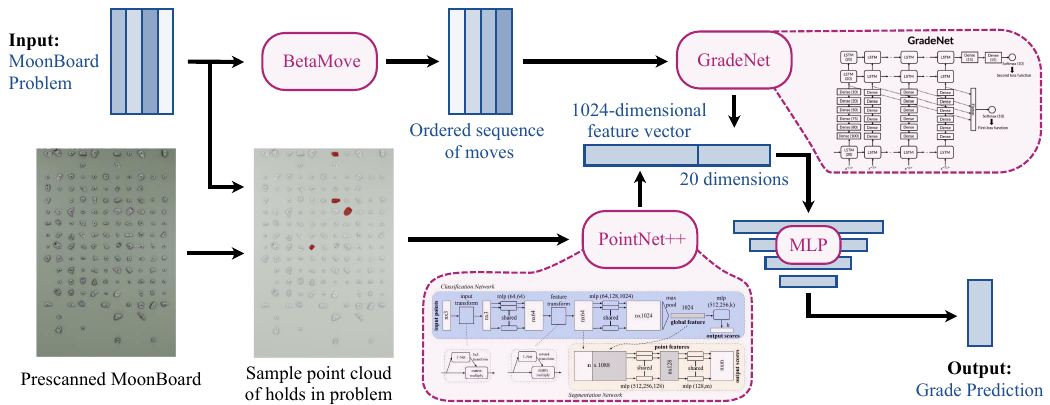}
 \centering
  \caption{A MoonBoard problem is a climbing route that consists only of a subset of holds from a standardized MoonBoard (bottom left). From a dataset of scanned holds, our neural algorithm uses this geometric knowledge to predict the grade of each specific problem.}
\label{fig:teaser}
}

\maketitle
\begin{abstract}
    A \emph{MoonBoard} is a standardized bouldering wall used in gyms around the world.
    Climbs up the wall limited to only a subset of holds are known as \emph{problems}.
    We introduce \emph{PointGrade}, a novel machine learning approach to predicting the difficulty of a MoonBoard \emph{problem}. 
    By sampling a point cloud from pre-scanned meshes of every hold, our model combines 3D object classification architecture with existing sequence-based approaches to difficulty grade prediction.
    Our method captures latent geometric information contained the climb, outperforming other work on the problem that neglect this data.
    \\

\end{abstract}  

\begin{table*}
\caption{We quantitatively evaluate the performance of our method on real MoonBoard problems uploaded by the climbing community. Our method outperforms all baselines (taken from \cite{petashvili2023board,duh2021recurrent}), including the prior state of the art GradeNet~\cite{duh2021recurrent}, on all test metrics.}
\begin{center}
\rowcolors{2}{gray!10}{white}
\begin{tabularx}{\textwidth}{
    >{\centering\arraybackslash}p{3.8cm} ||
    >{\centering\arraybackslash}p{2.5cm} |
    >{\centering\arraybackslash}X |
    >{\centering\arraybackslash}X |
    >{\centering\arraybackslash}p{2.5cm} |
    >{\centering\arraybackslash}X }
\rowcolor{header2color}
\hline
\toprule
 Method & Training Accuracy & Eval Accuracy  & Test Accuracy & Test Accuracy ±1 & Test F1 Score \\
\midrule
GradeNet
    &49.06\% & 48.61\% & 47.78\% & 86.98\% & 0.2347  \\ 
PointNet++ 
    &51.91\% & 51.21\% & 50.86\% & 84.18\% & 0.1877 \\
2DCNN 
    &60.00\% & 49.92\% & 48.14\% & 86.38\% & 0.2406 \\
Linear Regression 
    &36.36\% & 34.64\% & 37.20\% & 83.46\% & 0.1453 \\
Linear SVR
    &39.91\% & 38.66\% & 40.33\% & 82.90\% & 0.1526 \\
RBF SVR
    &62.92\% & 49.55\% & 48.70\% & 87.71\% & 0.2153 \\
Decision Tree Regressor
    &\textbf{95.41}\% & 41.32\% & 41.33\% & 77.21\% & 0.2251 \\
XGBoost Regressor
    &49.45\% & 43.41\% & 44.65\% & 85.70\% & 0.2016 \\
    \midrule
Ours (Concurrently Trained)
    & 52.38\% & 52.75\% & 51.41\% & 84.45\% & 0.1903 \\
Ours (No Hold Geometry)
    & 51.58\% & 52.83\% & 52.42\% & 84.96\% & 0.1900 \\
\textbf{Ours}
    & 55.48\% & \textbf{55.46}\% & \textbf{54.18}\% & \textbf{88.01}\% & \textbf{0.2750} \\
\bottomrule
\end{tabularx}
\end{center}
\label{tab:table}
\end{table*}

\section{Introduction}

The MoonBoard is an 11x18 grid of standardized rock climbing holds in a specific preset layout used in rock climbing gyms around the world (see \autoref{fig:teaser}). Climbers challenge themselves to scale this wall using only small subsets of the board's 139 holds; these subsets are known as \emph{problems}. The official MoonBoard app
hosts a large database of problems manually created by climbers, who subjectively assign a difficulty score to their problem when uploading. Without any formal guidelines for scoring a problem, problems with the same score can vary greatly in difficulty depending on the climber, creating a need for a more reliable method to grade them. 

Unfortunately, existing methods of assigning a score to a problem struggle to reach high levels of accuracy. Current state-of-the-art methods that train machine learning models to score a problem using only the identities of each hold \cite{duh2021recurrent,petashvili2023board} overlook the rest of the information contained in the problem, leading to lower accuracies than could otherwise be achieved. This task is further complicated by the recent introduction of ML-generated \cite{duh2021recurrent} problems that lack a human creator to assign them a grade. 

In this paper we introduce PointGrade, a geometric approach to scoring MoonBoard problems with machine learning. We began by manually scanning a set of MoonBoard holds, which allowed us to convert any problem into a detailed 3D model of the MoonBoard for that problem. Using a geometric deep learning model we project each problem into a latent space and incorporate these encodings into existing models trained on only the IDs of the holds in the problem. By combining the perspective of previous work with this additional geometric information, we predict grades that account for the sequence of holds as well as for which \emph{shapes} of holds are harder or easier to grasp in a given sequence. 

We experimentally validate every algorithmic decision, from point cloud size to the method of recreating the 3D model of the MoonBoard. By training and testing our model on a large pre-existing community dataset, we demonstrate that incorporating geometric information into these methods allows for a greater accuracy than can be achieved using the IDs of the holds alone. With these results, we not only set a new state of the art in scoring the difficulty of MoonBoard problems, but also draw insights into the geometric nature of this task that hopefully inspire future work.

\section{Related Work}

Our method builds upon existing work in MoonBoard problem difficulty prediction, a specialized area of study in the larger field of general climbing difficulty prediction. To situate our method in the context of previous work, we begin by reviewing the various techniques explored in classifying the difficulty of general climbs (\autoref{sec:general-climbing-related-work}), then focus in on machine learning approaches to predicting the difficulty of MoonBoard problems (\autoref{sec:moonboard-related-work}).

\subsection{General Climbing Difficulty Prediction}
\label{sec:general-climbing-related-work}

Within the sport of climbing, routes are assigned a \emph{grade} depending on their difficulty, decided by experts or in some cases the larger community of climbers. Methods of predicting these grades are commonly categorized into climber-centric approaches and route-centric approaches \cite{o2025addressing}.

\emph{Climber-centric approaches} record biometric data during climbs with tools like wearable sensors \cite{ladha2013climbax} then compare this data with responses from the climbers \cite{delignieres1993psychophysical}, with the goal of eventually predicting the difficulty of a bouldering problem based on biometric data alone \cite{ebert2017automated}. In some instances, these approaches will forego the biometric data, instead applying statistical models to large databases of self-reported climbs to predict a grade, such as the Whole-History Rating method proposed by Scarff \cite{scarff2020estimation}, later refined via  Bayesian Markov Chain Monte Carlo inference \cite{drummond2021bayesian}. This approach may introduce bias in the form of selective self-reporting from climbers \cite{drummond2021bayesian}, which can be reduced with automatic climb-logging from data gathered using tools such as wrist-worn intertia measurement units \cite{kosmalla2015climbsense}.

\emph{Route-centric approaches} analyze individual segments of the climb to draw conclusions about the grade of the entire climb, like the probabilistic method proposed by Ansel \cite{ansel2023probability} that calculates an energy cost associated with each of these segments to derive a probability distribution for the entire problem's grade. Many route-centric methods treat the climb as a natural language processing problem, parsing instructions for individual moves between holds into machine interpretable symbols \cite{phillips2012strange}. The variable-order Markov model \emph{StrangeBeta} was trained using these tokenized climbing instructions to generate novel climbs \cite{phillips2012strange}, and later adapted into a difficulty predictor that observes a route and classifies it as either "easy" or "hard" \cite{kempen2018fair}. This work was foundational in grade prediction for the MoonBoard, the focus of this paper, whose standardized layout and large database of problems are well suited for the machine learning approaches that we review in the next section.

\subsection{MoonBoard Problem Grade Classification}
\label{sec:moonboard-related-work}

The MoonBoard is an 11x18 grid of standardized rock climbing holds in a specific preset layout used in rock climbing gyms around the world \cite{moonboardwebsite}
When training a neural network to grade MoonBoard problems, the most common approach treats each problem as a one-hot encoding vector (where the position of each hold in the problem corresponds to an entry in the vector coded as "1"), proposed by Dobles et al. \cite{dobles2017machine} who train a convolutional neural network using these encodings. Recent work builds upon this foundation, preprocessing the data gathered from the MoonBoard website before training a 2D convolutional neural network to improve the quality of the data and consequentially achieve accuracies comparable to the state of the art in MoonBoard difficulty prediction \cite{petashvili2023board}.
Additional work has attempted to train graph graph neural networks \cite{tai2020graph}, long-short term memory (LSTM) networks \cite{houghton2020moon}, and other models for this task, however without the preprocessing step, these methods struggle to reach comparable accuracies. With no barriers to uploading a problem, MoonBoard's database includes many problems with inaccurately labeled grades, and across various methods of encoding and training, preprocessing this data has consistently been shown to improve accuracy.

Most relevant to our method is the work of Chang and Duh \cite{duh2021recurrent}, which first introduced the preprocessing step used in most MoonBoard problem grade classifcation models today.  The novel preprocessing pipeline \emph{BetaMove} first filters out problems with zero attempts and problems mislabeled as the highest difficulty,  then uses a beam sort algorithm to find the ideal sequence of holds to traverse a problem, optimized to match the hold sequence provided by a sample of expert climbers. Each move in this sequence is embedded into a 22-dimensional vector used to train their LSTM recurrent neural network \emph{GradeNet}, which achieves a 46.7\% test set accuracy, representing the current state of the art in grading MoonBoard problems.

In short, existing MoonBoard problem grade classifiers have largely been trained on the position and sequence of the holds with moderate success. Our method exploits the MoonBoard's standardized layout to capture geometric information implied in a problem, training a grade classification model using this data and the latent encodings from a modified verion of \emph{GradeNet} to achieve higher accuracies than previously reported (see \autoref{tab:table}).

\section{Method}

Our method uses machine learning to predict the difficulty of a given problem.
The input to our model is a single moonboard problem which, given the standardized layout, can be represented as a list of grid coordinates $(i_1,j_1),\dots,(i_k,j_k)$. Our output will be a \emph{grade}, given as an integer from $0$ to $9$, corresponding to the grades V4 to V13 along the commonly used \emph{Hueco scale} for grading MoonBoard problems, the scale upon which our dataset is graded.

\paragraph*{Dataset.}

\begin{figure}
    \centering
    \includegraphics[width=1\linewidth]{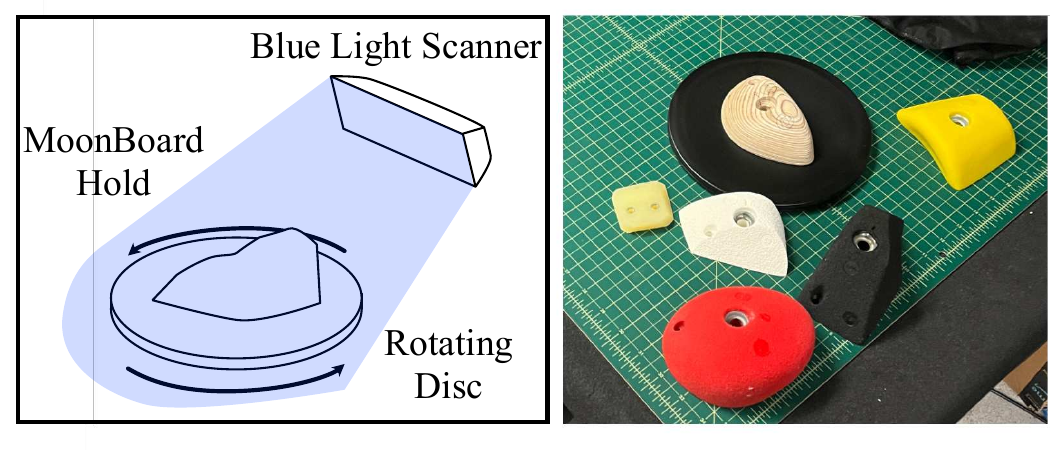}
    \vspace{-0.75cm}
    \caption{We scanned all holds using a Creality Raptor Pro 3D Scanner and a rotating disc.}
    \label{fig:scanning-setup}
\end{figure}

To train our model, we utilize a collection of meshes corresponding to every hold in the 2016 MoonBoard set, as well as a database of problems uploaded by the climbing community.
To build our collection of hold meshes, we purchased every hold in the 2016 MoonBoard layout and scanned each hold using the Creality Raptor Pro 3D scanner's blue laser mode, rotating the hold on a rotating disc as we scanned (see \autoref{fig:scanning-setup}). We reoriented each hold and repeated this process to capture every side of it, before combining the resulting point clouds into a single scan. Using the scanner's built-in software, we converted each point cloud into a high-resolution ($100$k-$300$k faces) triangle mesh, which we manually oriented to match the hold's orientation in the official MoonBoard layout. We use \textsc{Qslim} \cite{garland1997surface} as implemented in \textsc{Gpytoolbox} \cite{gpytoolbox} to decimate these into medium resolution meshes with $10,000$ triangles each.
Our dataset of problems is taken from Chang and Duh \cite{duh2021recurrent}, who extracted them from the official MoonBoard app and filtered them to remove ``low quality'' submissions, with a split of $20157$ training, $2442$ evaluation and $2497$ test problems.

\paragraph*{Architecture.}
Our goal is to combine the insights from the identity-based approaches with geometric priors to improve grading accuracy. 
To do this, our architecture will consist of two branches (see \autoref{fig:teaser}). One processes the MoonBoard problem into a sorted sequence of climbing moves using \emph{BetaMove} \cite{duh2021recurrent}, a beam-search algorithm, followed by the convolutional architecture proposed by \emph{GradeNet}, to produce a $20$-length feature vector.
Our other branch starts from the full scanned MoonBoard, discarding the meshes not involved in the current problem. The remaining hold meshes are sampled to produce a full $1024$-point point cloud, which is passed to \emph{PointNet++} \cite{qi2017pointnet++} to produce a $1024$-size feature vector. This feature vector is concatenated with the $20$-length one produced in the prior branch, and passed through a flattening MLP with layers $1044\rightarrow512\rightarrow 256 \rightarrow 10$ and ReLU+Dropout activations. The output is a vector of size 10; its highest entry is our output grade.

\paragraph*{Training.}
Our model contains weights for three learnable elements: GradeNet, PointNet++ and our final MLP. Initially, we considered training them all concurrently, which was enough to outperform all our considered baselines (see \autoref{tab:table}, Ours (Concurrently Trained) row); however, we devised a three-step strategy that leads to improved accuracy.
First, we trained the top branch of our model, including only the GradeNet weights, but swapping the final layer for a flattening to a $10$-dimension vector which encodes the problem's difficulty.
Once this branch has been trained to match the best possible difficulty (a process identical to the training proposed by Chang and Duh \cite{duh2021recurrent}), we train the PointNet++ branch on its own to similarly grade each problem.
Once both branches have been trained, their final layers are removed and their last hidden layer outputs are used as the concatenated feature vectors that are passed to our final MLP.
This MLP's weights are optimized in one last training loop. PointNet++ and the MLP are both trained using cross entropy loss, while GradeNet uses a weighted cross entropy loss to account for the skewed dataset. All trainings are performed using Adam \cite{kingma2014adam} optimization.

\section{Results}

We implemented our method in Python, building on the official implementations released by GradeNet \cite{duh2021recurrent} and PointNet++ \cite{qi2017pointnet++}. We performed all experiments on a Linux machine with an NVIDIA RTX A6000 GPU with 51.5 GB of memory and 1 TB of RAM.
All baseline comparisons are carried out using the implementations shared by Petashvili and Rodda \cite{petashvili2023board} and Chang and Duh \cite{duh2021recurrent}.

Our goal was to incorporate geometric information into the MoonBoard grading process to improve its accuracy. To accomplish this, we integrated PointNet++ into a standard sequence-based grading architecture. A critical question is whether the performance improvements we observed can be attributed to the new information conveyed by the shape of the hold, or some other property deduced by PointNet++ (e.g., the relative positioning of the holds). To ablate this, we carry out an experiment in which all holds are assumed to have the same geometry (in particular, the geometry of hold \#1 of the Original School Holds set). The results of this experiment are shown in \autoref{tab:table}: accuracy is substantially lower, justifying our conclusion that the shape of the hold significantly influences the difficulty of a specific climb.

\begin{figure}
    \centering
    \includegraphics[width=1\linewidth]{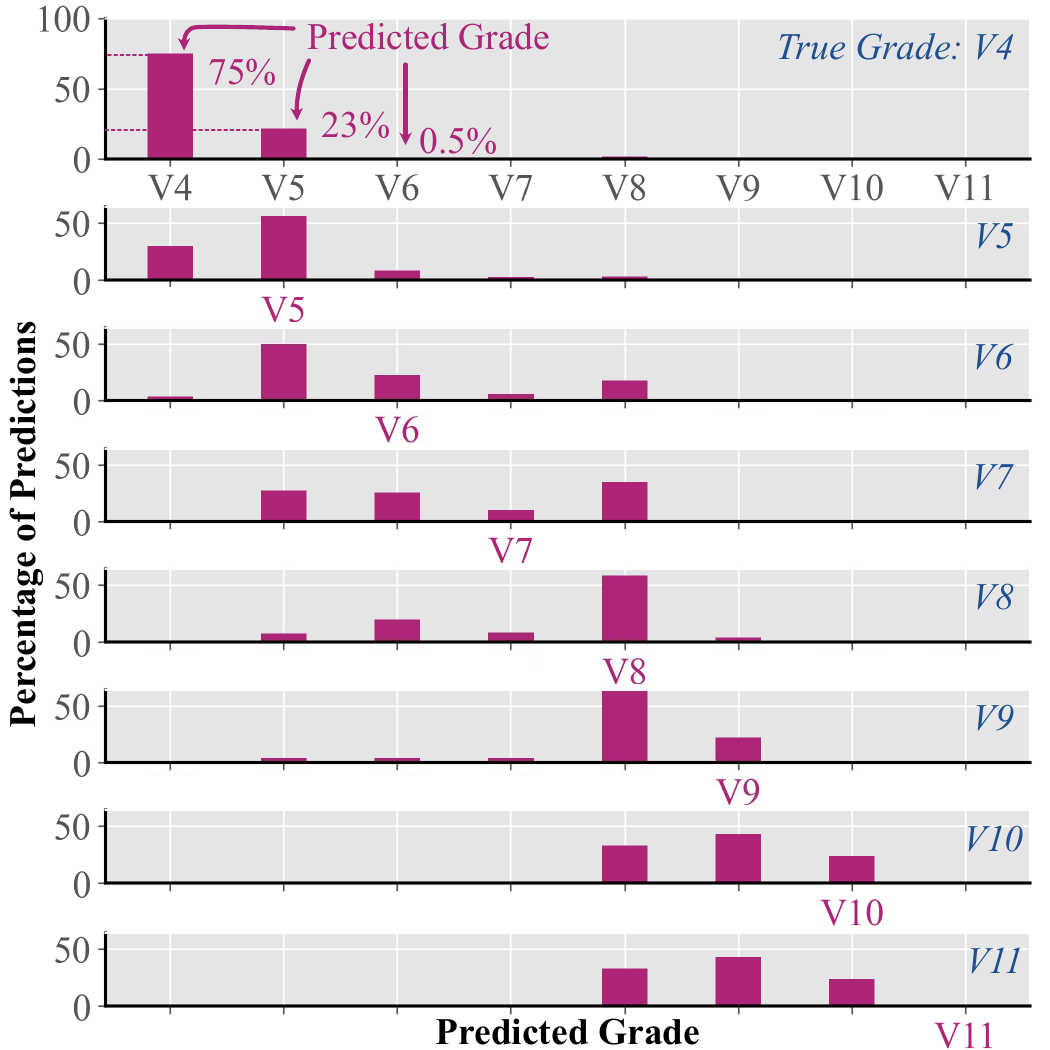}
    \vspace{-0.5cm}
    \caption{Our method is most often correct; when it fails, it usually does so by outputting a neighboring grade.}
    \label{fig:grade_preds}
\end{figure}

\begin{figure}
    \centering
    \includegraphics[width=1\linewidth]{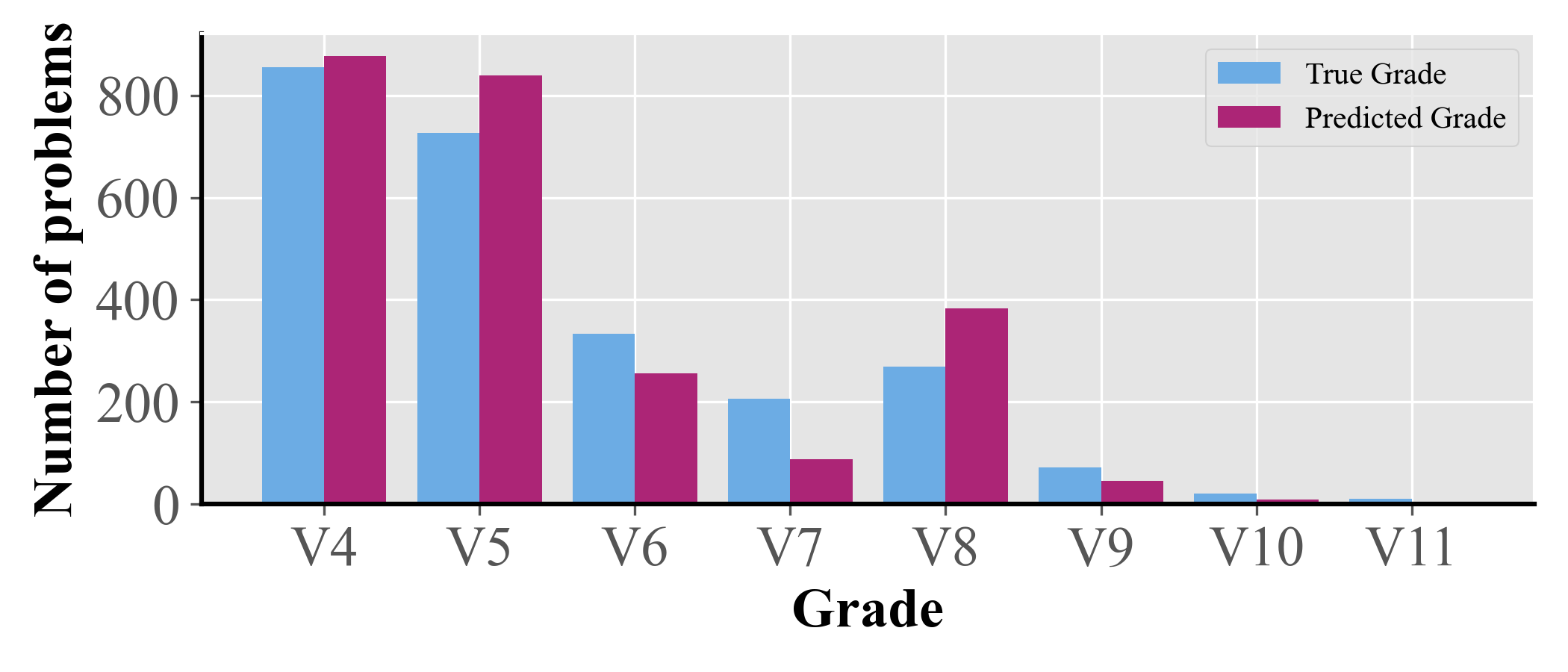}
    \vspace{-0.75cm}
    \caption{Our dataset is skewed towards lower grades, and our predictions mimic this concentration.}
    \label{fig:pred_vs_true}
\end{figure}

Because it contains a representative sample of problems uploaded by the climbing community, our dataset is skewed, concentrating around the lowest grades. As demonstrated in \autoref{fig:grade_preds} and \autoref{fig:pred_vs_true}, our method mirrors this concentration.
As shown in \autoref{fig:grade_preds}, when it fails, it often does so by suggesting a neighboring (most often lower) grade.

\begin{figure*}
    \centering
    \includegraphics[width=1\linewidth]{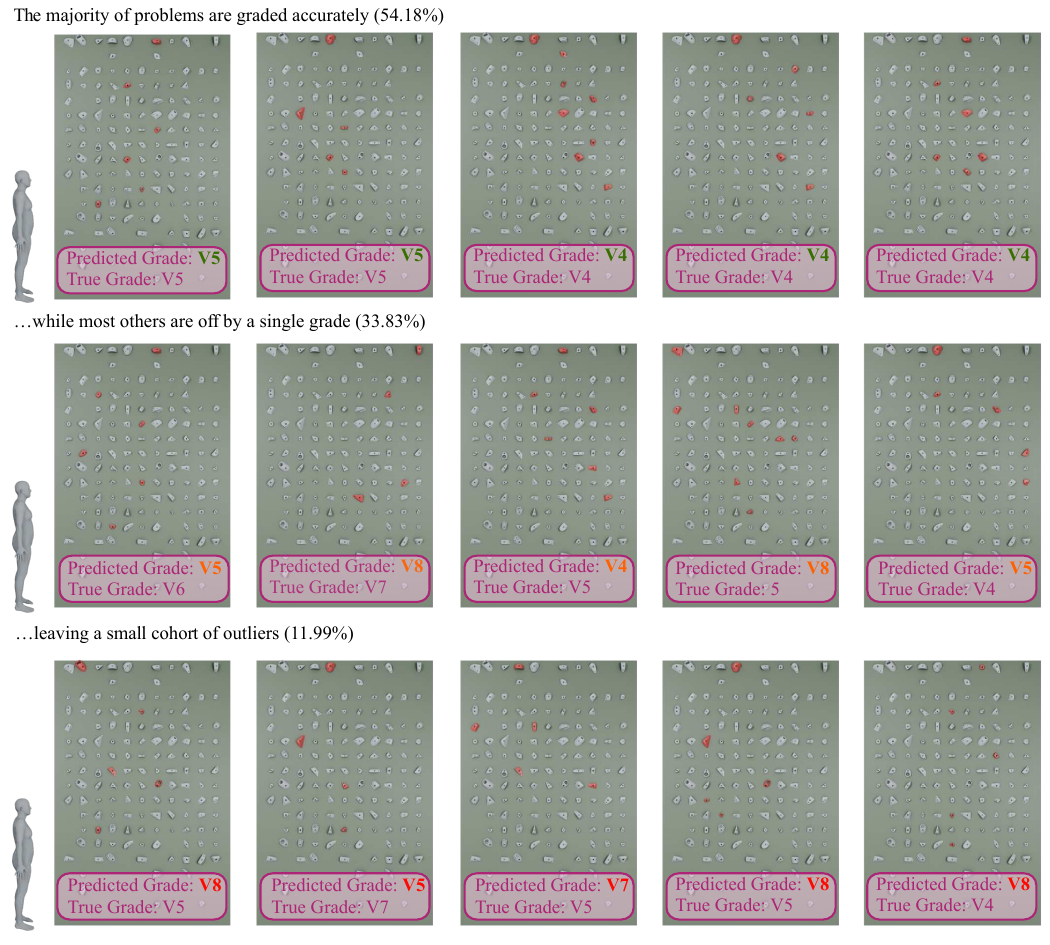}
    \vspace{-0.75cm}
    \caption{Qualitative the results shown in \autoref{tab:table}, showing examples in which our method succeeds, fails by one, or by more.}
    \label{fig:correct_close_wrong}
\end{figure*}

Our flagship result is a large-scale quantitative comparison (see \autoref{tab:table}) with a range of previously proposed baselines for the MoonBoard grading problem, including the prior state of the art. As suggested by Chang and Duh \cite{duh2021recurrent}, we report not only the exact test accuracy (the percentage of problems in our test set that each method grades correctly), but also the $\pm1$ accuracy; i.e., how many problems are graded correctly up to a difference of one.
Across all metrics, our method obtains the highest test accuracy and $\pm1$ test accuracy.
We qualitatively explore examples of cases in which our method succeeds (54.18\%), ``fails'' by one (33.83\%) and fails (11.99\%) in \autoref{fig:correct_close_wrong}.

\section{Conclusions and Limitations}

Our method introduces a novel approach to MoonBoard problem difficulty prediction, setting a new state of the art in the process.
We hope it can be of use to the climbing community as a standard for grading both community-uploaded and automatically generated problems.

To train our automated grading system to its reported level of accuracy, we rely on a dataset in which the labels (grades) have been subjectively assigned by the uploader of each problem. Without a standard metric for grading problems, climbs will differ in difficulty between climbers, leading to a ceiling in accuracy (and motivating the standard choice of reporting both exact and $\pm1$ accuracies).
In the future, we hope to augment our model to accept information about each climber as input and predict the grade that a specific climber would assign to the problem.

More broadly, our work highlights the importance of geometric information in predicting the difficulty of a climb. While this point is better demonstrated in the standardized setting of the MoonBoard, we hope our findings are useful in a broader set of applications, from recreational bouldering to robotic grasping and disaster rescue planning.

\section{Acknowledgements}

The Geometry and the City lab at Columbia University is supported by generous gifts from nTop, Adobe, Dandy and Braid Technologies, as well as by a research sponsorships from Meshy and a grant from the Columbia Engineering Interdisciplinary Research Fund.
This research project was sponsored by Dreamsports through the Columbia-Dream Sports AI Innovation Center. 

\bibliographystyle{eg-alpha-doi}
\bibliography{references}

\end{document}